\documentclass[12pt]{revtex4}  

\usepackage{amsmath}    
\usepackage{amsfonts}  
\usepackage{amssymb}
\usepackage{dsfont}
\usepackage{mathrsfs}
\usepackage{graphicx}   
\usepackage{txfonts}

\begin{document}

\newcommand{\PT}{$\mathcal{PT}$}
\newcommand{\vett}[1]{\mathbf{#1}}
\newcommand{\sech}{\text{sech}}

\title{Quasi $\mathcal{PT}$-symmetry in passive photonic lattices}

\author{Marco Ornigotti$^1$ and Alexander Szameit$^1$}
\address{$^1$Institute of Applied Physics, Friedrich-Schiller Universit\"at Jena, Max-Wien Platz 1, 07743 Jena, Germany}
\email{marco.ornigotti@uni-jena.de}

\begin{abstract}
The concept of quasi-\PT symmetry in optical wave guiding system is elaborated by comparing the evolution dynamics of a \PT-symmetric directional
coupler and a passive directional coupler. In particular we show that in the low loss regime, apart for an overall exponentially damping factor that
can be compensated via a dynamical renormalization of the power flow in the system along the propagation direction, the dynamics of the passive
coupler fully reproduce the one of the \PT-symmetric system.
\end{abstract}


\date{\today}
\maketitle

\section{Introduction}
Quantum mechanics is one of the brightest scientific theories of the last century. When one faces this theory for the first time, one of the
fundamental requirements is that the Hamiltonian operator associated to a dynamical system must be Hermitian, a necessary condition for its
eigenvalues to be real \cite{ref1}. In 1998, however, Bender and co workers \cite{ref2} demonstrated that this strict condition can be relaxed in
favor of a more weaker one, by introducing the concept of \PT-symmetry.  The so-called \PT-symmetric systems are characterized by a complex
potential, which per se possesses neither parity symmetry ($\mathcal{P}$) nor time-reversal symmetry ($\mathcal{T}$), but the Hamiltonians of these
systems share the same eigenstates with the parity-time operator \PT. Under these conditions, the eigenvalues of the Hamiltonian are real (up to a
certain threshold below which \PT-symmetry is preserved) even if that the potential is complex-valued.

Despite the interesting nature of these works, and the significant number of discussion on the impact of \PT-symmetry in quantum mechanics itself
\cite{ref4}, quantum field theory \cite{ref5}, open quantum systems \cite{ref6} and Anderson localization phenomena \cite{ref7}, this extension of
the ordinary quantum theories remained, however, a mere speculative curiosity mainly because of the difficulty of finding in Nature a system that
exhibits such a peculiar Hamiltonian.

Almost in parallel with the development of \PT-symmetric quantum mechanics, optics, and in particular waveguiding structures, attracted a huge
interest as a laboratory tool for mimicking the evolution dynamics of quantum systems, thanks to the mathematical correspondence between the paraxial
wave equation that describes the propagation of light in a guiding structure and the Schr$\mathrm{\ddot{o}}$dinger equation \cite{ref9}.  The
analogy, moreover, does not limit itself only to the non relativistic case, and very recently optical analogues of relativistic effects such as Klein
tunneling \cite{ref10}, Zitterbewegung \cite{ref12} and pair production in vacuum \cite{ref13} have been proposed. In 2007, the concept of
\PT-symmetry was brought into optics \cite{ref14}, revealing that optical systems are the natural candidates to realize in an easy way \PT-symmetric
systems. In the last years, \PT-symmetric optical systems attracted a large amount of interest and this topic was extensively studied \cite{ref17,
ref18, ref19, ref20, ref21,ref21a,ref21b}.

A common feature of all these \PT-symmetric optical systems is that in order to achieve \PT-symmetry, an equal amount of gain and losses must be
present in the system. However, in several works \emph{quasi}-\PT-symmetric systems were considered, in which only loss is present in the individual
waveguides and no gain \cite{ref16,Markus,Toni}. Such a scheme of an optical system partly lossless and partly lossy was also studied well before the introduction of \PT-symmetry as a way to reduce soliton jitter and noise power in optical fibers \cite{malomed}. It is the aim of this paper to provide a detailed proof that in such passive systems the evolution
dynamics is in fact the same as in systems with exhibiting gain and loss, up to a global exponential damping factor and for sufficiently small
losses. We base our considerations on the prototypical example of a \PT-symmetric directional coupler, but the extension to general lattices is
straightforward.

This work is organized as follows: in Sect. 2 we briefly review the salient features of a \PT-symmetric directional coupler with balanced gain and
losses. in Sect. 3 we introduce the concept of quasi-\PT-symmetry by analyzing the evolution of a light field in a directional coupler with
unbalanced losses and we show that the dynamical features are exactly the same as in the full-\PT-case. Conclusions are then drawn in Sect. 4.

\section{\PT-symmetric optical coupler}

\begin{figure}[!t]
\begin{center}
\includegraphics[width=\textwidth]{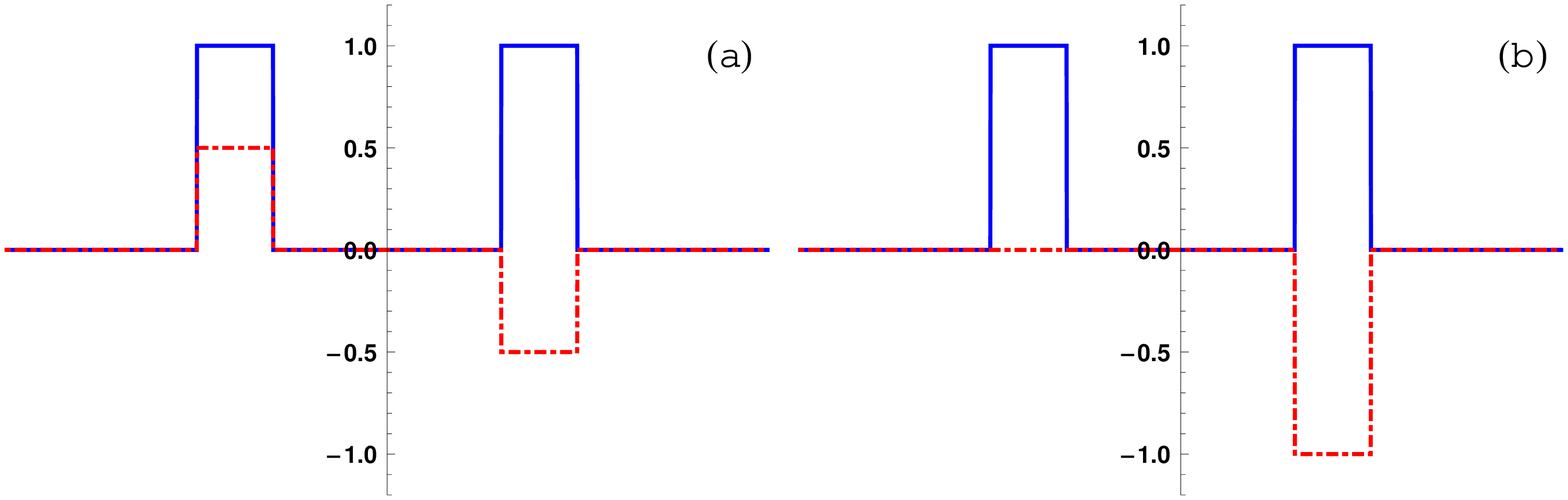}
\caption{Schematic representation of the complex refractive index profile $V(x)$ for the full \PT case [(a)] and the quasi-\PT case [(b)]. The real part of the potential $V_R(x)$ (blue solid line)  is the effective index profile of the coupler, while the imaginary part $V_I(x)$ (red dot-dashed line) accounts for gain and losses. Note that while in panel (a) $V_I(x)$ is anti-symmetric with respect to the geometrical center of the coupler (reflecting the full-\PT symmetry of the coupler, where the left waveguide experiences gain and the right one an equal amount of losses), in panel (b) this is not true anymore, as the quasi-\PT symmetry condition implies that only one waveguide would experience losses, whose absolute value is double with respect to the case depicted in (a).}
\label{figure1}
\end{center}
\end{figure}

We begin our analysis by considering a monochromatic scalar electric field $\psi(x,z)$ characterized by a frequency $\omega_0$ and a wavelength
$\lambda=2\pi/k$ that propagates inside a directional coupler. We assume that the mode field inside the coupler is strongly confined along the
$y$-direction, so that the beam dynamics in such a system can be taken to be one-dimensional in the transverse plane. The evolution of the light
field $\psi(x,z)$ inside such a structure is governed by the following dimensional paraxial equation:
\begin{equation}\label{eq1}
i\frac{\partial\psi}{\partial z}=-\frac{\partial^2\psi}{\partial x^2}+V(x)\psi\equiv\hat{\mathcal{H}}\psi,
\end{equation}

where $z$ and $x$ are suitably chosen dimensional coordinates and $V(x)=V_R(x)+iV_I(x)$ is the complex optical potential that implements \PT-symmetry
in the system. Following Ref. \cite{ref14},  in order for $V(x)$ to be \PT-symmetric the condition $V^*(x)=V(-x)$ must be fulfilled. This condition
arises from the necessity for the Hamiltonian in Eq. \eqref{eq1} to commute with the parity-time operator $\hat{\mathcal{PT}}$, in such a way that
$\psi$ is a common eigenstate of both $\hat{\mathcal{H}}$ and $\hat{\mathcal{PT}}$. The parity operator $\hat{\mathcal{P}}$ is defined by the
operations $\hat{x}\rightarrow -\hat{x}$ and $\hat{p}\rightarrow -\hat{p}$, while the time reversal operator $\hat{\mathcal{T}}$ consists of the
operations $\hat{p}\rightarrow -\hat{p}$ and $i\rightarrow -i$ \cite{ref3}. Practically, this means that the optical complex potential $V(x)$
consists of a symmetric refractive index profile $V_R(x)$ and an antisymmetric gain/loss profile $V_I(x)$A sketch of the refractive index profile of such a coupler is given in Fig. \ref{figure1} (a). By expanding the scalar electric field
$\psi(x,z)$ onto the eigenmodes of the coupler as
\begin{equation}
\psi(x,z)=[a_1(z)u_1(x)+a_2(z)u_2(x)]e^{i\beta z},
\end{equation}
where $\beta$ is the real (due to \PT-symmetry) propagation constant and the eigenmodes $u_k(x)$ are normalized according to \cite{ref14,ref14ter}
\begin{equation}
\int dx\; u_m^*(-x)\;u_n(x)=\delta_{mn},
\end{equation}
\begin{figure}[!t]
\begin{center}
\includegraphics[width=\textwidth]{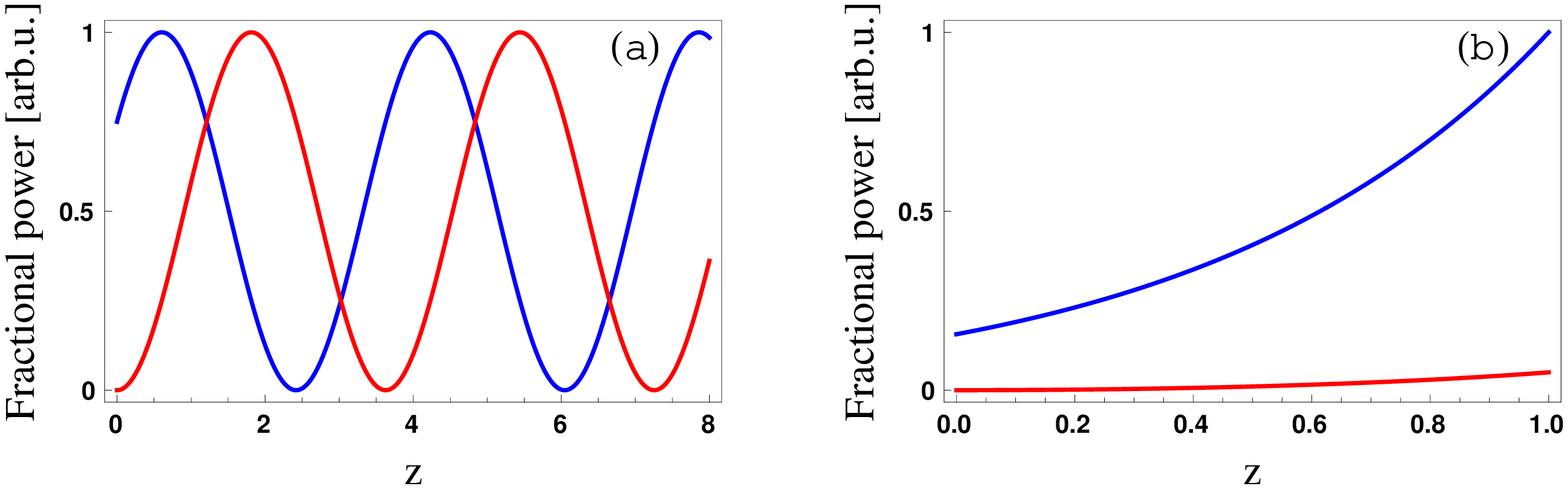}
\caption{(a) Normalized fractional powers $|a_1(z)|^2$  (blue line) and $|a_2(z)|^2$ (red line)  as a function of the normalized distance $z$ below
the \PT-symmetry breaking threshold for $\kappa/\gamma=2$. The evolution shows the characteristic non reciprocal power transfer. (b) Normalized
fractional powers $|a_1(z)|^2$  (blue line) and $|a_2(z)|^2$ (red line)  as a function of the normalized distance $z/$ above the \PT-symmetry
breaking threshold for $\kappa/\gamma=1/2$. The curves are normalized to the maximum of the fractional power contained in the first waveguide. As can
be seen, power grows exponentially faster with propagation. } \label{fig2}
\end{center}
\end{figure}
we can describe the evolution of light in such a system via the following coupled mode equations:
\begin{equation}\label{ptcmt}
i\frac{d}{dz}\left(
\begin{array}{c}
a_1\\ a_2
\end{array}
\right)= \left(
\begin{array}{cc}
\Delta+i\gamma & \kappa \\ \kappa & \Delta-i\gamma \\
\end{array}
\right) \left(
\begin{array}{c}
a_1\\  a_2
\end{array} \right),
\end{equation}
where $\kappa$ is the usual coupling coefficient and $\Delta+i\gamma$ is the shift in the propagation constant due to the coupling interaction. Note
that here instead of following the design of Ref. \cite{ref14} where a single waveguide experiences both gain and losses in equal amount, we employ
the dimer idea developed in Ref. \cite{ref14bis}, where \PT-symmetry is obtained by inserting gain in the first waveguide and losses in the second
one in equal amount. We note moreover that the real part $\Delta$ can be neglected because it can be eliminated by a gauge transformation, that
corresponds to consider the two waveguides to have no relative detuning \cite{ref22}. Without loss of generality we then set $\Delta=0$ in Eq.
\eqref{ptcmt}. According to Ref. \cite{ref16}, as long as $\kappa/\gamma>1$ the \PT-symmetry is unbroken and light is periodically exchanged between
the two waveguides. On the other hand, \PT-symmetry is said to be broken when $\kappa/\gamma<1$ and the light dynamics become exponentially growing
in one waveguide and exponentially damping in the other one \cite{ref16}. Light evolution in these two regimes is depicted in Fig. \ref{fig2}.

\section{Quasi-\PT-symmetric optical coupler}
The directional coupler described in the previous section fully implements  a \PT-symmetric system, and it is realized according to the rule
described in Ref. \cite{ref14bis}, namely to insert gain in one waveguide and losses in the other one in equal measure. Although this appears to be
the normal way of building a \PT-symmetric structure in optics, it appears to be quite complicated to be realized experimentally, as a full control
of gain and losses is a very crucial task. It is therefore interesting to study whether a similar physical problem as the one described in the
previous section can be obtained by exploiting only passive systems, and create a loss unbalance between the two waveguides instead of a gain and
loss structure. Let us then consider a directional coupler in which each waveguide experiences a different level of losses, and no gain is inserted
in the system. The complex refractive index profile is delicted in Fig. \ref{figure1} (b). Using a standard coupled-mode theory \cite{ref22}, the propagation of a light beam in such a structure can be described as follows:
\begin{equation}\label{cmt}
i\frac{d}{dz}\left(
\begin{array}{c}
a_1\\ a_2
\end{array}
\right)= \left(
\begin{array}{cc}
-i\gamma_1 & \kappa \\ \kappa & -i\gamma_2 \\
\end{array}
\right) \left(
\begin{array}{c}
a_1\\  a_2
\end{array} \right),
\end{equation}
where $\gamma_{1,2}$ accounts for the losses in the first and second waveguide respectively and $\kappa$ has been defined before as the coupling
coefficient. It is interesting to compare this equation with Eq. \eqref{ptcmt} with $\Delta=0$. While in eq. \eqref{ptcmt} there is only a sign
difference between the two diagonal elements, here in principle $\gamma_1\neq\gamma_2$ but the sign is the same. The sign discrepancy is due to the
fact that while Eq. \eqref{ptcmt} describes a gainy/lossy system, Eq. \eqref{cmt} describes a lossy system only. Note, moreover, that since the
diagonal elements of the matrix in Eq. \eqref{cmt} are purely imaginary, they cannot be removed via a simple phase transformation. We can however
exploit a common trick used in quantum field theory known as Wick rotation \cite{ref23}, that consist in rotating the time axis (in this case the
propagation axis) by $\pi/2$ in the complex plane, thus employing a complex time (in our case a complex propagation direction) instead of a real one.
This trick is very useful especially in lattice quantum field theory, to transform the Minkowski metric to an euclidean metric, allowing methods of
statistical mechanics to be used for evaluating path integrals on a lattice \cite{ref24}.  If we define $\zeta=i z$ and substitute this ansatz into
Eq. \eqref{cmt}, we obtain

\begin{figure}[!t]
\begin{center}
\includegraphics[width=\textwidth]{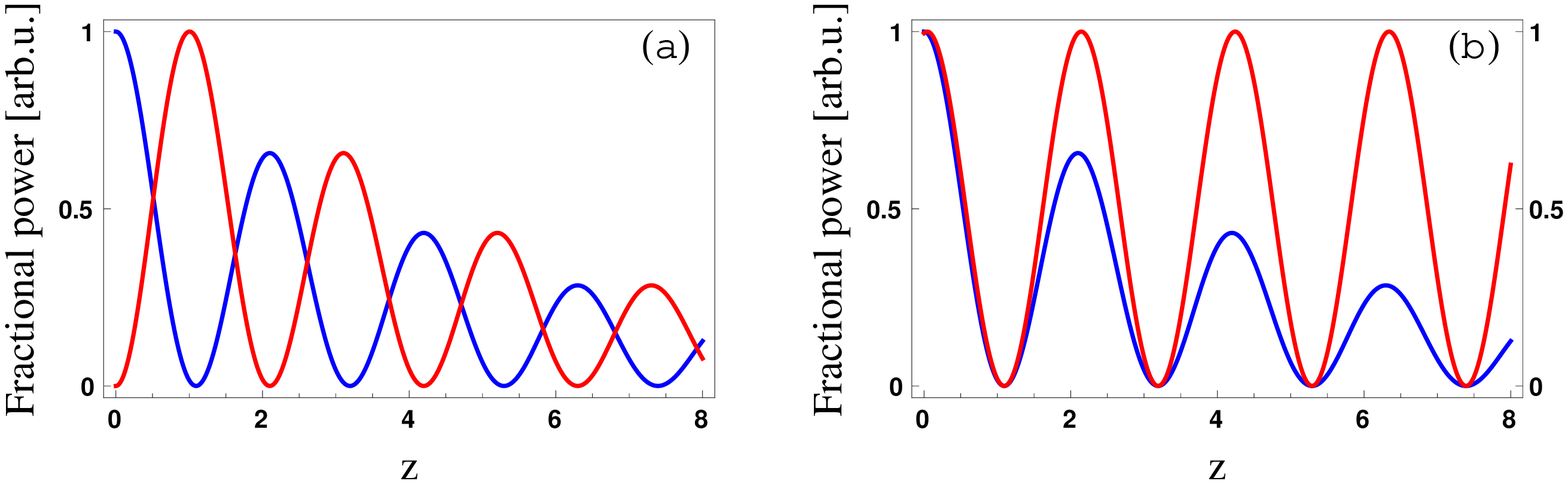}
\caption{(a) Evolution of the fractional powers $|\tilde{a}_1(z)|^2$ (blue line) and $|\tilde{a}_2(z)|^2$ (red line) for the unbroken quasi-\PT-symmetry with
$\kappa/\gamma=15$. (b) Comparison between the evolution  in the unbroken \PT-symmetry regime ($\kappa/\gamma=15$) of the fractional power for the
quasi-\PT-symmetric system $|\tilde{a}_1(z)|^2$  (blue line) and the corespondent \PT-symmetric system $|c_1(z)|^2$ (red line). As can be seen, apart from an
exponential damping factor that progressively reduces the intensity along $z$, the dynamics in the two cases is identical.} \label{fig3}
\end{center}
\end{figure}

\begin{equation}\label{wick}
-\frac{d}{d\zeta}\left(
\begin{array}{c}
a_1\\ a_2
\end{array}
\right)= \left(
\begin{array}{cc}
-i\gamma_1 & \kappa \\ \kappa & -i\gamma_2 \\
\end{array}
\right) \left(
\begin{array}{c}
a_1\\  a_2
\end{array} \right).
\end{equation}
We can now make the following phase transformation
\begin{equation}\label{phase}
\left(\begin{array}{c}
a_1\\ a_2
\end{array}
\right)= \left(\begin{array}{c} \tilde{a}_1\\ \tilde{a}_2
\end{array}
\right)e^{i\gamma_a\zeta},
\end{equation}
and then transform back to the real propagation axis $z$ to obtain
\begin{equation}\label{cmt2}
i\frac{d}{dz}\left(
\begin{array}{c}
\tilde{a}_1\\ \tilde{a}_2
\end{array}
\right)= \left(
\begin{array}{cc}
0 & \kappa \\ \kappa & -i(\gamma_2-\gamma_1) \\
\end{array}
\right) \left(
\begin{array}{c}
\tilde{a}_1\\ \tilde{a}_2
\end{array} \right).
\end{equation}
Before proceeding with the analysis of Eq. \eqref{cmt2}, a couple of words of explanations on this procedure are needed. First, note that thanks to
Eq. \eqref{phase}, to restore the initial amplitudes $a_{1,2}(z)$ one needs to multiply the solutions of Eq. \eqref{cmt2} by an exponentially damping
factor $\exp{\left(-\gamma_1 z\right)}$. While theoretically this only accounts to a gauge transformation in Wick space, experimentally the presence
of this extra damping term is not a problem since once the amount of losses $\gamma_1$ is known, this term can be easily eliminated via post
processing of the acquired image.

It is now instructive to calculate the eigenvalues of the matrix that appears in Eq. \eqref{cmt2} and compare them with the ones from Eq.
\eqref{ptcmt}. If we call $\mu_{1,2}$ the eigenvalues of the \PT-symmetric system \eqref{ptcmt} and $\lambda_{1,2}$ the ones for the passive system
\eqref{cmt2}, we have the following result:

\begin{subequations}
\begin{align}
\mu_{1,2}=&\;\pm\sqrt{\kappa^2-\gamma^2},\\
\lambda_{1,2}=&\;-i\left(\frac{\gamma_2-\gamma_1}{2}\right)\pm\sqrt{\kappa^2-\left(\frac{\gamma_2-\gamma_1}{2}\right)^2}.
\end{align}
\end{subequations}
If we now choose $\gamma_2-\gamma_1=2\gamma$, where $\gamma$ is the same  value of gain/loss that appears in the \PT-symmetric case, then, apart from
a common imaginary part (that will result in an exponential damping factor), the dynamics of the passive system is the \PT-symmetric case, as
$\lambda_{1,2}=i\gamma+\mu_{1,2}$. This is the main result of this paper: the dynamics of a \PT-symmetric system are exactly the same as the dynamics
of a passive one (i.e., a system with only losses) provided that the losses of the system are chosen in such a way that $\gamma_2-\gamma_1=2\gamma$.

\begin{figure}[!t]
\begin{center}
\includegraphics[width=\textwidth]{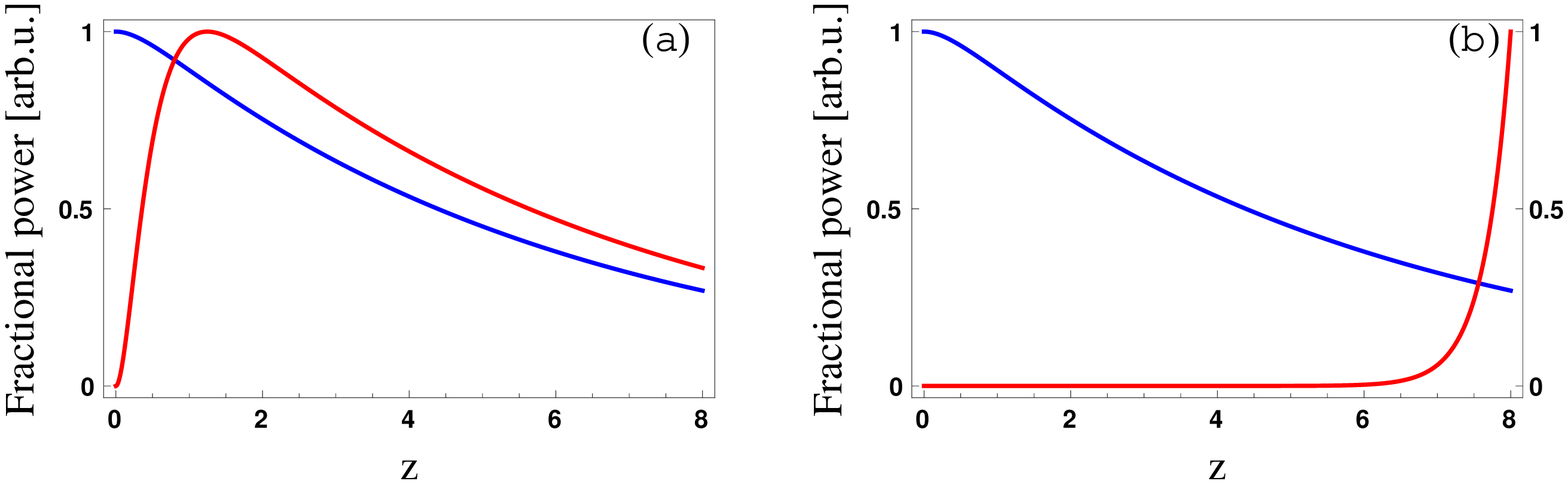}
\caption{a) Evolution of the fractional powers $|\tilde{a}_1(z)|^2$ (blue line) and $|\tilde{a}_2(z)|^2$ (red line) for the broken quasi-\PT-symmetry with $\kappa/\gamma=1/3$. (b) Comparison between the evolution  in the broken \PT-symmetry regime ($\kappa/\gamma=1/3$)  of the fractional power for the quasi-\PT-symmetric system $|\tilde{a}_1(z)|^2$  (blue line) and the corespondent \PT-symmetric system $|c_1(z)|^2$ (red line). As can be seen, above the \PT-threshold, the presence of the overall damping exponential term makes the dynamics of the quasi-\PT system very different from its \PT-counterpart. However, if the exponentially damping term is compensated by a $z$-dependent power normalization, the two dynamics perfectly coincide.}
\label{fig4}
\end{center}
\end{figure}

\begin{figure}[!t]
\begin{center}
\includegraphics[width=\textwidth]{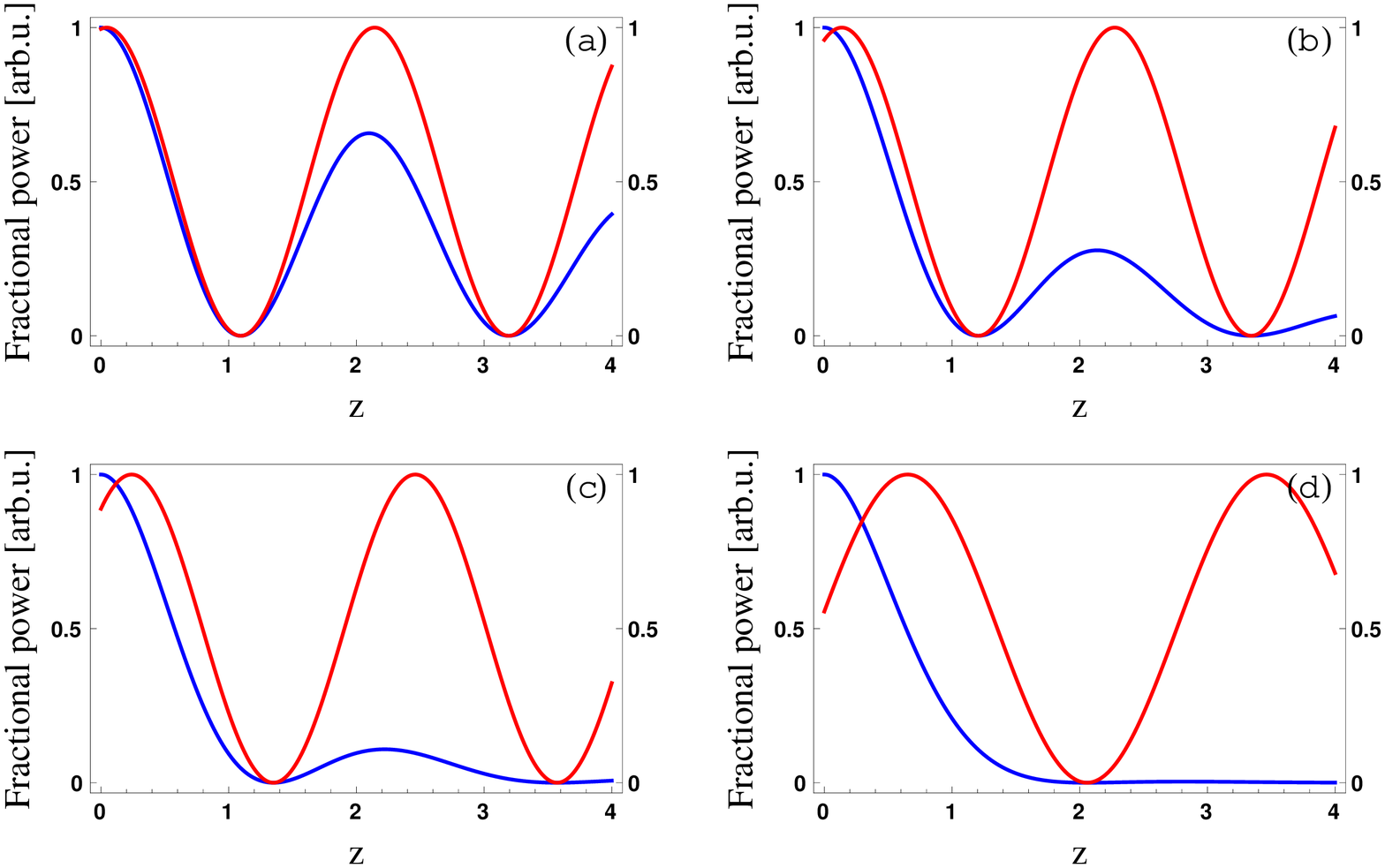}
\caption{Comparison of the dynamics of the evolution of the fractional power contained in the first waveguide for the quasi-\PT system $|\tilde{a}_1(z)|^2$
(blue line) and its \PT-counterpart $|c_1(z)|^2$ (red line) for different values of the losses in the system. For all these graphs, the value of the
coupling coefficient is $\kappa=1.5$. (a) $\gamma=0.1$, (b) $\gamma=0.3$, (c) $\gamma=0.5$ and (d) $\gamma=1$. As can be seen, as the losses
increase, the discrepancy between the two curves also increases, and the main effect on the dynamics is that the quasi-\PT system is no longer able
to faithfully reproduce the dynamics of its \PT-counterpart. This is in particular visible in panels (c) and (d) where the period of the two curves
is not matched.} \label{fig5}
\end{center}
\end{figure}

A comparison between the dynamics of a passive system as described by Eq. \eqref{cmt2} and the correspondent \PT-case is depicted in Fig. \ref{fig3}
for unbroken \PT-symmetry and in Fig. \ref{fig4} for the broken \PT-symmetry case. This is true for the dynamics of the passive system in terms of
the amplitudes $\tilde{a}_{1,2}(z)$. However, while below threshold (Fig. \ref{fig3}(b)) the presence of the exponential damping factor only affects the
intensity of the light that propagate in the quasi-\PT system, above threshold (Fig. \ref{fig4}(b)), where the \PT-symmetry is broken, the behavior
of the quasi-\PT system and the \PT one are profoundly different, and one needs to compensate for the overall damping factor dynamically by
performing a $z$-dependent normalization of the power evolution inside the system in order to restore the real \PT dynamics.
As a last remark, in Fig. \ref{fig5} the comparison between the quasi-\PT and the \PT- dynamics for different values of the ratio $\kappa/\gamma$
below threshold is reported. As it can be seen, the equivalence between the dynamics is only trivial when $\kappa/\gamma\gg1$ or $\gamma\ll 1$. In
this case, a compensation of the losses in the system is not needed, as the amount of losses is small enough not to disturb the underlying \PT-like
dynamics. If the losses increase, however, these dynamics start to get obscured by the high losses of the system, and an immediate correspondence
between the two cases cannot be established anymore. This is also consistent with the fact that in systems with high losses the light trapped in the
waveguides gets quickly absorbed or scattered away, and the characteristic length upon which the dynamics takes place is too small to allow any
interesting dynamics to be seen.

\section{Conclusions}
In conclusion, we have shown that a gain/loss structure as the one described in Ref. \cite{ref14bis} is not a necessary condition for an optical
system to show \PT-symmetry. By employing a passive (i.e., lossy) system, we proved that although this latter system follows a non-Hermitian
dynamics, for small enough losses the dynamical behavior of such a system is, up to an overall exponential damping factor, perfectly reproduces the
characteristic dynamics of a \PT-system. To prove this we performed a Wick rotation of the system and applied a gauge transformation in Wick space to
eliminate the losses in one of the waveguides and thus define the loss unbalance, that is the governing parameter of the system. Our results show
that below the \PT-symmetry breaking threshold, the dynamics of the two systems are fully equivalent (in the low loss regime), while above this
threshold, a dynamical power renormalization as a function of the propagation distance is needed in order to extract the fully \PT-dynamics.

As a final remark we note that our results can be of high importance for experimental realizations of \PT-symmetric optical systems, as controlling
the amount of losses in a passive system is surely an easier task than inserting a gain structure in an optical system.

\section*{Aknowledgements}
The authors thank the German Ministry of Education and Science (ZIK
03Z1HN31) for financial support.

\end{document}